\documentclass[aps,floatfix,nofootinbib,preprintnumbers,showpacs,prd,twocolumn,superscriptaddress]{revtex4-2}

\usepackage{overpic}
\usepackage{amssymb}
\usepackage{indentfirst}
\usepackage{feynmf}   
\usepackage{epsfig}
\usepackage{array}
\usepackage{subfigure}
\usepackage{graphicx}
\usepackage{lipsum}
\usepackage{slashed}
\usepackage{amsmath}
\usepackage{bm}
\usepackage{amssymb}
\usepackage{color,xcolor}
\usepackage[colorlinks,
            citecolor=green,
            anchorcolor=red,
            menucolor=red,
            linkcolor=red,
            filecolor=red,
            runcolor=red,
            urlcolor=blue,
            frenchlinks=red]{hyperref}

\begin{document}

\title{Gluon TMDs from J/$\psi$ production in longitudinally polarized deeply inelastic scattering}

\author{Huachao Liu}\affiliation{School of Physics, Southeast University, Nanjing
211189, China}
\author{Xiupeng Xie}\affiliation{School of Physics, Southeast University, Nanjing
211189, China}
\author{Zhun Lu}
\email{zhunlu@seu.edu.cn}
\affiliation{School of Physics, Southeast University, Nanjing 211189, China}

\begin{abstract}

We investigate the feasibility of exploring the gluon transverse-momentum-dependent distribution functions (TMDs) inside a longitudinally polarized nucleon.
We utilize quarkonium production via the color-octet mechanism combined with TMD formalism in semi-inclusive deeply inelastic scattering (SIDIS) at low transverse momentum as a tool to access polarized gluon TMDs.
The corresponding cross-section of the process is expressed in terms of gluon TMDs and non-relativistic QCD matrix elements.
We provide the expressions for the $\sin2\phi$ azimuthal asymmetries of $J/\psi$ production in SIDIS with a unpolarized beam colliding on longitudinally polarized nucleon. The asymmetry is contributed by the time-reversal-odd gluon TMD $h_{1L}^{\perp \,g}(x,\boldsymbol{p}_T^2)$.
The maximum possible asymmetry deduced from the positivity bound is sizable and could be measured.
We also estimate the double longitudinal spin asymmetry $A_{LL}$ of $J/\psi$ production using a spectator model result for $g_{1L}^g(x,\bm p_T^2)$.
\end{abstract}

\maketitle

\section{introduction}

The study of the 3-dimensional partonic structure of nucleons has gained significant interests~\cite{Mulders:1995dh,Boer:1997nt,Mulders:2000sh,Goeke:2005hb,Bacchetta:2006tn}.
The corresponding knowledge is encoded in the transverse momentum dependent parton distribution functions (TMDs) of quarks and gluon. 
TMDs contain not only the information of the light-cone momentum fraction $x$, but also the transverse motion $\bm p_T$
 of partons inside a nucleon. 
Certain TMDs can be aroused from the correlation between spin and partonic transverse momenta, and give rise to single-spin asymmetries~\cite{Sivers:1989cc,Anselmino:1994tv,Brodsky:2002cx,Airapetian:2010ds,Qian:2011py,Adolph:2014zba} in high-energy process.
Furthermore, TMDs reflects nontrivial QCD dynamics due to the gauge-link structure~\cite{Bomhof:2006dp} of their operator definitions, which leads to their process dependence, such as the sign change of time-reversal-odd (T-odd) TMDs between SIDIS and Drell-Yan processes~\cite{Collins:2002kn}. 

Recently the quark TMDs have been extensively studied by theoretical approaches as well as experimental measurement.
On the contrary, the gluon TMDs~\cite{Mulders:2000sh} are almost unknown from experimental aspects.
This is because the processes sensitive to gluon require high energy collision and gluon TMDs are difficult to disentangle from the contributions of quark TMDs. 
Theoretically and phenomenologically, several gluon TMDs were explored in literature. 
The unpolarized gluon TMD $f_1^g(x,\bm p_T^2)$ and the linearly polarized gluon distribution inside an unpolarized nucleon $h_1^{\perp\,g}(x,\bm p_T^2)$ have been calculated in the context of the small $x$ physics~\cite{Dominguez:2010xd,Dominguez:2011br}.
Proposals to measure $h_1^{\perp\,g}(x,\bm p_T^2)$ have been suggested  through jet~\cite{Boer:2010zf,Dumitru:2015gaa} or heavy quark pair production~\cite{Boer:2010zf,Pisano:2013cya} in ep/pp collision and heavy quarkonia~\cite{Boer:2012bt,denDunnen:2014kjo} production.
Meanwhile, the unpolarized TMD gluon distribution has been extracted~\cite{Lansberg:2017dzg} from LHCb data on the transverse
spectra of $J/\psi$ pairs by assuming a Gaussian shape.
Regarding the naive time-reversal-odd (T-odd) gluon TMDs, the gluon Sivers function~\cite{Sivers:1989cc,Boer:2015vso}  has been studied by model calculations~\cite{Goeke:2006ef,Lu:2016vqu,Yao:2018vcg} and
phenomenological studies~\cite{DAlesio:2015fwo,DAlesio:2017rzj,Rajesh:2018qks,Zheng:2018ssm}.

In Refs.~\cite{Boer:2016fqd}, the process $e p\rightarrow e^\prime Q \bar{Q} X$ has been explored to access the gluon TMDs  $f_{1T}^{\perp g}$, $h_1^{\perp,g}$, $h_1^g$ and $h_{1T}^{\perp g}$  at an Electron-Ion Collider~(EIC). 
Different azimuthal asymmetries have been defined to single out certain gluon TMD.
Recently, the feasibility to access gluon TMDs via azimuthal asymmetries in the processes of $J/\psi$~\cite{Bacchetta:2018ivt,Chakrabarti:2022rjr} or $\Upsilon$~\cite{Bacchetta:2018ivt} production off unpolarized or transversely polarized target has been explored.
Apart from giving access to the gluon TMDs $f_{1T}^{\perp g}$, $h_1^{\perp,g}$, $h_1^g$ and $h_{1T}^{\perp g}$, the study also provides a new method to extract specific color-octet NRQCD long-distance matrix elements.
In this work, we present a comprehensive analysis on the  $ e+p \rightarrow e^{\prime} J/\psi X $ process with the electron or the proton target either longitudinally polarized or unpolarized. 
Similar to Ref.~\cite{Bacchetta:2018ivt}, in this analysis we adopt TMD framework in combination with nonrelativistic QCD (NRQCD)~\cite{Hagler:2000dd,Yuan:2000qe,Yuan:2008vn}.
It is notable that color-octet diagrams and color-singlet diagrams contribute to quarkonia differential cross-sections at the same order~\cite{Cho:1995vh}. 
A quarkonium is primarily the interaction between the heavy quark $Q$ and its antiquark $\bar{Q}$ in a bound state. 
In the initial analysis, the color-octet amplitude was hypothesized to have two parts: a short-distance component and a long-distance component that contains all the non-perturbative QCD effects. The long-distance component was measured in non-relativistic wave function or its derivatives. The primary benefit of NRQCD is the capacity to differentiate between various order contributions.
The production cross-sections can be calculated systematically to any preferred order in both the strong coupling constant and the velocity~\cite{Bodwin:1994jh,Braaten:1996jt}. 

Specifically, we provide the expressions for the azimuthal asymmetry of $J/\psi$ production in SIDIS with a unpolarized beam colliding on longitudinally polarized nucleon. 
The T-odd parton distributions on TMDs are significant in giving rise to the azimuthal asymmetries~\cite{Sivers:1989cc}. The gluon TMD $h_{1L}^{\perp\,g}$, often referred to as the gluon Kotzinian-Mulders function, is a vital element in this context. 
$h_{1L}^{\perp\,g}$ denotes the linearly polarization of the gluon inside a longitudinally proton and has seldom been explored in literature.
It gives rise to the $\sin2\phi_T$ azimuthal asymmetry in the process, with $\phi_T$ the azimuthal angle between  the transverse momentum of the $J/\psi$ and the lepton plane.
We also estimate the double longitudinal spin asymmetry $A_{LL}$ of $J/\psi$ production using a spectator model result for $g_{1L}^g$~\cite{Bacchetta:2020vty}.

The rest of the paper is organized as follows. In Section.~II, we present the operator definition of the gluon TMDs and introduce the spectator model. In Section.~III, a QCD analysis of the production of a heavy quarkonium is presented, including the derivation of the amplitude of the S-wave and P-wave states in the partonic subprocess. 
In Section.~IV the cross-section of $J/\psi$ production in polarized SIDIS is derived by incorporating the contribution of the antisymmetric component of the leptonic tensor. In Section.~V, we conduct a numerical study that utilizes the positivity bound to evaluate the $\sin2\phi_T$ azimuthal  asymmetry in the kinematic range accessible at EIC. We summarize the work in Section.~VI.

\section{Gluon TMDs}
	
In order to connect the partons involved in a hard scattering process to the corresponding hadrons in the final or initial state, it is customary to use parton-parton correlation functions as the soft part of the process~\cite{Mulders:1995dh,Boer:1997nt,Mulders:2000sh,Goeke:2005hb,Bacchetta:2006tn}.
In particular, the gluon TMDs can be defined by using the following gluon-gluon correlator within a non-light-like interval. Since only the soft parts integrated over the momentum component $k^{-}$ are considered in this process, one can define:
\begin{align}\label{eq:correkationfunction}
    \Gamma^{\mu \nu}(x, \boldsymbol{p}_{T})=&\int d\xi^{-} d^2 \xi_{T}e^{ip\cdot \xi} \langle P,S |F^{+ \mu}(0) \mathcal{U} (0,\xi) \nonumber\\
    &\times  F^{+ \nu}(\xi)|P,S\rangle |_{\xi^{+}=0} \,.
\end{align}
The variables in this equation are denoted as follows: $P$ and $S$ are the momentum and spin of the hadron, respectively, whereas $p$ is the momentum of the gluon.

The gluon-gluon correlator depends on the momentum fraction $x$, the transverse momentum $\boldsymbol{p}_{T}$, as well as the momentum and spin of the target.
It is convenient to project the correlator on the basis of transverse tensors and vectors and parameterize it using gluon TMDs~\cite{Mulders:2000sh}.
The correlator can be further categorized based on the type of hadron spin dependency: unpolarized (O), longitudinally polarized (L), and transversely polarized (T).
This leads to:
\begin{align}\label{eq:TMDO}
	\Gamma_{O}^{\mu \nu}=&\frac{x}{2}\left[-g_{T}^{\mu \nu}f_1^g\left(x, \boldsymbol{p}_{T}^{2}\right)\right. \notag \\
	& \left.+\left\{\frac{p_{T}^{\mu}p_{T}^{\nu}}{M_{p}^{2}}+g_{T}^{\mu \nu}\frac{ \boldsymbol{p}_{T}^{2}}{2M_{p}^{2}}\right\}h_1^{\perp \,g}\left(x, \boldsymbol{p}_{T}^{2}\right)\right]\,,
\end{align}
in leading twist, where $f_1^g\left(x, \boldsymbol{p}_{T}^{2}\right)$ is the TMD unpolarized distribution, and $h_1^{\perp\,g}\left(x, \boldsymbol{p}_{T}^{2}\right)$ is the distribution of linearly polarized gluons.
And
\begin{align}\label{eq:TMDL}
    \Gamma_{L}^{\mu \nu}=&\frac{x}{2}\left[-i\epsilon_{T}^{\mu\nu}S_{L} g_{1L}^g\left(x, \boldsymbol{p}_{T}^{2}\right)\right. \notag \\
	&\left. +\frac{\epsilon_{T}^{p_{T}\{i}p_{T}^{j\}}}{2M_{p}^2}S_{L}h_{1L}^{\perp\, g}\left(x, \boldsymbol{p}_{T}^{2}\right)\right]\,.
\end{align}
Here, $ g_{1L}\left(x, \boldsymbol{p}_{T}^{2}\right)$ represents the difference of the numbers of gluons with opposite circular polarization in a longitudinally polarized nucleon; $h_{1L}^{\perp\, g}\left(x, \boldsymbol{p}_{T}^{2}\right)$
is the gluon Kotzinian-Mulders function representing the linearly polarization of the gluon in a longitudinally polarized nucleon. Finally, in the transverse basis, the projection reads
\begin{align}\label{eq:TMDT}
    \Gamma_{T}^{\mu \nu}=&\frac{x}{2}\left[-g_{T}^{\mu \nu}\frac{\epsilon_{T}^{p_{T} S_{T}}}{M_{p}}f_{1T}^{\perp\,g}\left(x, \boldsymbol{p}_{T}^2\right)-i\epsilon_{T}^{\mu\nu}
     \frac{ \boldsymbol{p}_{T}\cdot \boldsymbol{S}_{T}}{M_{p}}g_{1T}^g\left(x, \boldsymbol{p}_{T}^2\right)\right. \nonumber \\
     &+\frac{\epsilon_{T}^{p_{T}\{\mu}p_{T}^{\nu\}}}{2M_{p}^2}\frac{ \boldsymbol{p}_{T}\cdot  \boldsymbol{S}_{T}}{M_{p}}h_{1T}^{\perp\,g}\left(x, \boldsymbol{p}_{T}^2\right) \nonumber\\ &+\frac{\epsilon_{T}^{p_{T}\{\mu}S_{T}^{\nu\}}+\epsilon_{T}^{S_{T}\{\mu}p_{T}^{\nu\}}}{4M_{p}}\left[ h_{1T}^g\left(x, \boldsymbol{p}_{T}^2\right)\right.\nonumber\\
     &\left.\left.-\frac{ \boldsymbol{p}_{T}^2}{2M_{p}^2} h_{1T}^{\perp\,g}\left(x, \boldsymbol{p}_{T}^2\right)\right]\right]\,,
\end{align}
where $g_{1T}^g\left(x, \boldsymbol{p}_{T}^{2}\right)$ represents the difference of the numbers of gluons with opposite circular polarization in a longitudinally polarized nucleon, $f_{1T}^{\perp\,g}\left(x, \boldsymbol{p}_{T}^2\right)$ is the gluon sivers function, while $ h_{1T}^g\left(x, \boldsymbol{p}_{T}^2\right)$ and $h_{1T}^{\perp\,g}\left(x, \boldsymbol{p}_{T}^2\right)$ are chiral-even distributions of linearly polarized gluons inside a transversely polarized nucleon. 
Among the eight leading-twist gluon TMDs,  $f_{1T}^{\perp\,g}$, $h_{1L}^{\perp\, g}$,  $ h_{1T}^g$ and $h_{1T}^{\perp\,g}$ are T-odd.

In our calculation the involved gluon TMDs are $f_1^g(x,\bm p_T^2)$, $h_{1L}^{\perp\, g}$ and $g_{1L}^{ g}$. 
In order to predict relevant observables for gluon TMDs, we adopt the spectator model results from Ref.~\cite{Bacchetta:2020vty}. 
The model is based on the assumption that a nucleon can emit a gluon, and that what remains after the emission is treated as a single spectator particle. The model expression for a generic gluon TMD $\Gamma^{g}\left(x, \boldsymbol{p}_{T}^2\right)$ weighing on the spectral function $\rho_{X}$ reads:
\begin{align}\label{eq:model_expression_of_TMD}
	F^{g}(x, \boldsymbol{p}_{T}^2)=\int_{Mp}^{\infty} d M_{X} \rho_{X}\left(M_{X}\right) \hat{F}^{g}\left(x, \boldsymbol{p}_{T}^2;M_{X}\right)\,.
\end{align}

Apart form the model results , model independent constraints such as the positivity bounds can be applied on the gluon TMDs. For $g_{1L}^g\left(x, \boldsymbol{p}_{T}^2\right)$ and $h_{1L}^{\perp\,g}\left(x, \boldsymbol{p}_{T}^2\right)$, the following bounds~\cite{Bacchetta:1999kz,Mulders:2000sh} 
\begin{align}\label{eq:PositiveBound1}
	\left| g_{1L}^g\left(x, \boldsymbol{p}_{T}^2\right)\right| &\le f_1^g\left(x, \boldsymbol{p}_{T}^2\right)\,.
 \\
	\frac{ \boldsymbol{p}_{T}^2}{2M_{p}^2}\left|h_{1L}^{\perp\,g}\left(x, \boldsymbol{p}_{T}^2\right)\right| &\le f_1^g\left(x, \boldsymbol{p}_{T}^2\right)\, \label{eq:sharpPositiveBound3}.
\end{align}
can be derived from the positivity of the corresponding matrix elements for the distribution functions in the gluon spin $\otimes$ nucleon spin space established in Ref.~\cite{Mulders:2000sh}.

\section{$J/\psi$ production in NRQCD framework}	

We consider the electroproduction of $J/\psi$ in longitudinally polarized SIDIS by including the color-octet contributions from the various partial wave states  $(c \bar{c})_{8}(^{2S+1} L_{J})$.
At the partonic level, the relevant subprocess is $\gamma^{*} (q)+ g(p) \rightarrow J/\psi(P_{\mathcal{Q}}) $, where the state $(c \bar{c})_{8}(^{2S+1} L_{J})$ evolves into a physical $J/\psi$ by absorbing soft gluons at the long distance scale~\cite{Ko:1996xw}.
As pointed out in Ref.~\cite{Fleming:1997fq}, the color-singlet contribution is relatively suppressed by a perturbative coefficient of order $ \alpha_{s}/\pi$ compared to the color-octet one.
Within the framework of NRQCD, the scattering amplitudes at leading order for partonic processes $\gamma^{*} (q)+ g(p) \rightarrow c \bar{c} (P_{\mathcal{Q}}) $ contributing to $ e+p \rightarrow e^{\prime} J/\psi X $ can be obtained using the approaches in Refs.~\cite{Bacchetta:2018ivt,Boer:2012bt}, while the long distance parts are treated as parameters $\langle  0|\mathcal{O}_{8}^{J / \psi} (^{2S+1}L_{J})|0\rangle $. 
The amplitudes of parton processes can be evaluated from~\cite{Bacchetta:2018ivt}:
\begin{widetext}
	\begin{align}
		\mathcal{A}^{\mu \nu}[^1 S_{0}^{(8)}](q,p)&=\frac{1}{4} \sqrt{\frac{\mathcal{C}_{S,0}}{2 M_{\mathcal{Q}}}} Tr [O^{\mu \nu}(0)(\slashed{P}_{\mathcal{Q}}-M_{\mathcal{Q}})\gamma^5]\,, \label{A1S0}\\
		\mathcal{A}^{\mu \nu}[^3 S_{1}^{(8)}](q,p)&=\frac{1}{4} \sqrt{\frac{\mathcal{C}_{S,1}}{2 M_{\mathcal{Q}}}} Tr [O^{\mu \nu}(0)(\slashed{P}_{\mathcal{Q}}-M_{\mathcal{Q}})\slashed{\varepsilon}_{s_{z}}] \,,\label{A3S1} \\
		\mathcal{A}^{\mu \nu}[^1 P_{1}^{(8)}](q,p)&=-i \sqrt{\frac{3\mathcal{C}_{P,0}}{8 N_{c} M_{\mathcal{Q}}}} Tr [(O^{\mu \nu}(0) \slashed{\varepsilon}_{L_{z}}\frac{\slashed{P}_{\mathcal{Q}}}{M_{\mathcal{Q}}}+\varepsilon_{L_{z}}^{\alpha}O_{\alpha}^{\mu \nu}(0)\frac{\slashed{P}_{\mathcal{Q}}-M_{\mathcal{Q}}}{2})\gamma^5]\,,\label{A1P1}\\
		\mathcal{A}^{\mu \nu}[^3 P_{0}^{(8)}](q,p)&=\frac{i}{2} \sqrt{\frac{\mathcal{C}_{P,0}}{2 N_{c} M_{\mathcal{Q}}}} Tr [-3 O^{\mu \nu}(0) + (\gamma^{\alpha} O_{\alpha}^{\mu \nu}(0)-\frac{\slashed{P}_{\mathcal{Q}} P_{\mathcal{Q}}^{\alpha} O_{\alpha}^{\mu \nu }(0)}{M_{\mathcal{Q}}^2})\frac{\slashed{P}_{\mathcal{Q}}-M_{\mathcal{Q}}}{2}]\,,\label{A3P0}\\
		\mathcal{A}^{\mu \nu}[^3 P_{1}^{(8)}](q,p)&=\frac{1}{4} \sqrt{\frac{3\mathcal{C}_{P,1}}{ N_{c} M_{\mathcal{Q}}}} \epsilon_{\alpha \beta \rho \sigma} \frac{P_{\mathcal{Q}}^{\rho}}{M_{\mathcal{Q}}} \varepsilon_{J_{z}}^{\sigma} (P_{\mathcal{Q}}) Tr [\gamma^{\alpha} O^{\beta \mu \nu} (0) \frac{\slashed{P}_{\mathcal{Q}}-M_{\mathcal{Q}}}{2} + O^{\mu \nu} (0) \frac{\slashed{P}_{\mathcal{Q}}}{M_{\mathcal{Q}}} \gamma^{\alpha} \gamma^{\beta}]\,,\label{A3P1}\\
		\mathcal{A}^{\mu \nu}[^3 P_{2}^{(8)}](q,p)&=i \sqrt{\frac{3\mathcal{C}_{P,2}}{8 N_{c} M_{\mathcal{Q}}}} \varepsilon_{J_{z}}^{\alpha \beta} (P_{\mathcal{Q}}) Tr[\gamma_{\alpha} O_{\beta}^{\mu \nu} (0)  \frac{\slashed{P}_{\mathcal{Q}}-M_{\mathcal{Q}}}{2}]\,,
	\end{align}
\end{widetext}
where $O^{\mu \nu}(0)\equiv O^{\mu \nu}(q,p,0,P_{\mathcal{Q}})$ and $O^{\mu \nu}_{\alpha}(0)\equiv O^{\mu \nu}_{\alpha}(q,p,0,P_{\mathcal{Q}})$ with
\begin{align}
O^{\mu \nu}(q,p,k,P_{\mathcal{Q}})&= -\sqrt{2} \delta^{ab} \frac{e e_{c} g_{s}}{2(M_{\mathcal{Q}}^2+Q^2)} \nonumber\\ &\times\left[\gamma^{\mu}(\slashed{p}-\slashed{q}+2\slashed{k}+M_{\mathcal{Q}})\gamma^{\nu}\right.\nonumber\\
&\left.-\gamma^{\nu}(\slashed{p}-\slashed{q}-2\slashed{k}-M_{\mathcal{Q}})\gamma^{\mu}\right]\,.\\
O^{\mu \nu}_{\alpha}&=\frac{\partial}{\partial k^{\alpha}}O^{\mu \nu}(q,p,k,P_{\mathcal{Q}})|_{k=0} \,,
\end{align}
 calculated at leading order in perturbative QCD.
Here, $k$ is half of the relative momentum of the outgoing quark-antiquark pair and we consider the SU(3) color-octet projector
We use the shorthand notation
$$\mathcal{C}_{S,J}=\frac{1}{2J+1} \langle
0|\mathcal{O}_{8}^{J / \psi} (^{2S+1}L_{J})|0\rangle $$

By employing substitutions, we can perform a summation over the $J/\psi$ spin states, some useful relations for polarization vectors can be adopted~\cite{Guberina:1980dc}.
Using $\varepsilon_{J_{z}}^{\mu}$ to denote the polarization vector for a bound state with total angular momentum $J=1$, the following polarization sum is satisfied
\begin{align}\label{eq:polarizationrelation1}
	\sum_{J_{z}=-1}^{1} \varepsilon_{J_{z}}^{\mu}\varepsilon_{J_{z}}^{*\nu}=-g^{\mu \nu} +\frac{P_{\mathcal{Q}}^{\mu} P_{\mathcal{Q}}^{\nu}}{M_{\mathcal{Q}}^2} \equiv \mathcal{P}^{\mu \nu}\,,
\end{align}
with $P_{\mathcal{Q}}$ the four-momentum of the state and $M_{\mathcal{Q}}$ the mass.
Finally, for a $J=2$ system, the summation of the polarization tensors $\mathcal{E}_{J_{z}}^{\mu \nu}$ yields
\begin{align}\label{eq:polarizationrelation2}
	\sum_{J_{z}=-2}^{2} \varepsilon_{J_{z}}^{\mu \nu}\varepsilon_{J_{z}}^{*\alpha \beta}=\frac{1}{2} [\mathcal{P}^{\mu \alpha}\mathcal{P}^{\nu \beta}+\mathcal{P}^{\mu \beta}\mathcal{P}^{\nu \alpha}]-\frac{1}{3}\mathcal{P}^{\mu \nu}\mathcal{P}^{\alpha \beta} \,.
\end{align}

\section{Differential cross-section for $J/\psi$ production in longitudinally polarized SIDIS}
    	
We will use the light-cone coordinate system, in which a vector $a^{\mu}$ is expressed as $a^{\mu}=(a^{+},a^{-},\vec{a}_{T})$. By means of the symmetric and antisymmetric transverse projectors, we introduce
\begin{align}
g_{T}^{\mu \nu}&=g^{\mu \nu}-\frac{P^{\mu}n^{\nu}}{P\cdot n}-\frac{P^{\nu}n^{\mu}}{P\cdot n}\,,\\ 	
\epsilon_{T}^{\mu \nu}&=\frac{\epsilon^{\alpha \beta \mu \nu}P_{\alpha}n_{\beta}}{P \cdot n}\,.
\end{align}    	

We choose the reference frame such that both the virtual photon exchanged in the reaction and the incoming proton move along the $\hat{z}-$ axis. Azimuthal angle $\phi_T$ is defined respect to the lepton scattering plane, such that $\phi_{\ell}=\phi_{\ell^{\prime}}=0$. The calculation follows the same lines of Ref.~\cite{Bacchetta:2018ivt}. The light-like vectors $n_{+}$ and $n_{-}$ satisfy the relation $n_{+}^2=n_{-}^2=0$ and $n_{+}\cdot n_{-}=1$, so we can write the four-momenta $P$ and $q$ as
\begin{align}\label{eq:momentumP}
    P&=n_{+}+\frac{M_{p}^2}{2}n_{-}\approx n_{+}\,,\\
    \label{eq:momentumq}
    q&=-x_{B}n_{+}+\frac{Q^2}{2 x_{B}}n_{-}\approx -x_{B}n_{+}+(P\cdot q)n_{-}\,,
\end{align}
where $Q^2=-q^2$ and $x_{B}$ is Bjorken variable,$M_{p}$ is the mass of the proton. 
The four-momenta of the leptons are expanded in terms of $P=n_{+}$ and $n=n_{-}=(q+x_{B}P)/P\cdot q$ by performing a Sudakov decomposition:
\begin{align}\label{eq:momentumell}
\ell &= \frac{1-y}{y}x_{B}P+\frac{1}{y}\frac{Q^2}{2x_{B}}n+\frac{\sqrt{1-y}}{y}Q\hat{\ell}_{\perp}\,,\\
\label{eq:momentumell'}
\ell^{\prime}&= \frac{1}{y}x_{B}P+\frac{1-y}{y}\frac{Q^2}{2x_{B}}n+\frac{\sqrt{1-y}}{y}Q\hat{\ell}_{\perp}\,,
\end{align}
where $y$ denotes the inelasticity variable $y=P\cdot q/P \cdot \ell$.
The invariant mass of the electron-target system is $s=(P+\ell)^2=2P \cdot \ell$, while the virtual photon-target invariant mass square is defined as $W^2=(P+q)^2=Q^2(1-x_{B})/x_{B}$. 
For the four momentum of the initial gluon $p$ and the final quarkonium state $\mathcal{Q}$ we have
\begin{align}
p&=x P +p_{T}=(p\cdot P -x M_{p}^2 )n\approx x P +p_{T}\,, \label{eq:momentump}\\
P_{\mathcal{Q}}&=z(P\cdot q)n +\frac{M_{\mathcal{Q}}^2+ \boldsymbol{P}_{\mathcal{Q}T}^2}{2z P\cdot q}P +P_{\mathcal{Q}T}\,,
\end{align}   	
with $z=P_{\mathcal{Q}}\cdot P/q \cdot P$ and $P_{\mathcal{Q}T}^2=- \boldsymbol{P}_{\mathcal{Q}T}^2$ .
    	    	
After introducing the matrix elements $\Gamma_{g}$ of the correlator for the gluon field strengths, the differential cross section can be written as
\begin{widetext}
    	\begin{equation}\label{sectiontheoretical}
    		d\sigma=\frac{1}{2s}\frac{d^3\ell'}{(2\pi)^3 2 E_{\ell}'}\frac{d^3 P_{\mathcal{Q}}}{(2\pi)^3 2E_{\mathcal{Q}}}\int dxd^2 \boldsymbol{p}_{T}(2\pi)^4 \delta^4 (q+p-P_{\mathcal{Q}})\frac{1}{x^2 Q^4}L_{\mu \rho}(\ell,q)\Gamma_{g \nu \sigma}(x, \boldsymbol{p}_{T})H_{\gamma*g\rightarrow \mathcal{Q}}^{\mu\nu}H_{\gamma*g\rightarrow \mathcal{Q}}^{*\rho\sigma}\,.
\end{equation}
\end{widetext}
The function $H^{\mu \nu}$ represents the scattering amplitude of the $\gamma^* g\rightarrow J/\psi$  process. 
The momentum conservation delta function can be expressed as
\begin{align}
    (2\pi)^4 \delta^4 (q+p-P_{\mathcal{Q}})  =&\frac{2}{ys}\delta\left(x-x_{B}-\frac{M_{\mathcal{Q}}^2+ \boldsymbol{P}_{\mathcal{Q}T}^2}{yzs}\right)\nonumber\\
   &\times \delta (1-z) \delta^2 ( \boldsymbol{p}_{T}- \boldsymbol{P}_{\mathcal{Q}T})\,, \label{eq:deltafunction}
\end{align}  	
corresponding to unpolarized and longitudinally polarized lepton beams, respectively. The leptonic tensor $L_{\mu \nu}$ can be decomposed into a symmetric and an antisymmetric part under $\mu \leftrightarrow \nu$ interchange.
Therefore, the leptonic tensor $L(\ell,q)$ is given by:
\begin{align}\label{eq:leptensor}
    L^{\mu \nu}(\ell,q)=e^2\left[-g^{\mu\nu}Q^2+2\left(\ell^{\mu}\ell'^{nu}+\ell^{\nu}\ell'^{\mu}\right)\right]	+2ie^2\epsilon^{\mu\nu \ell \ell'}\,.
\end{align}

\begin{figure*}
	\centering
		\includegraphics[width=80mm]{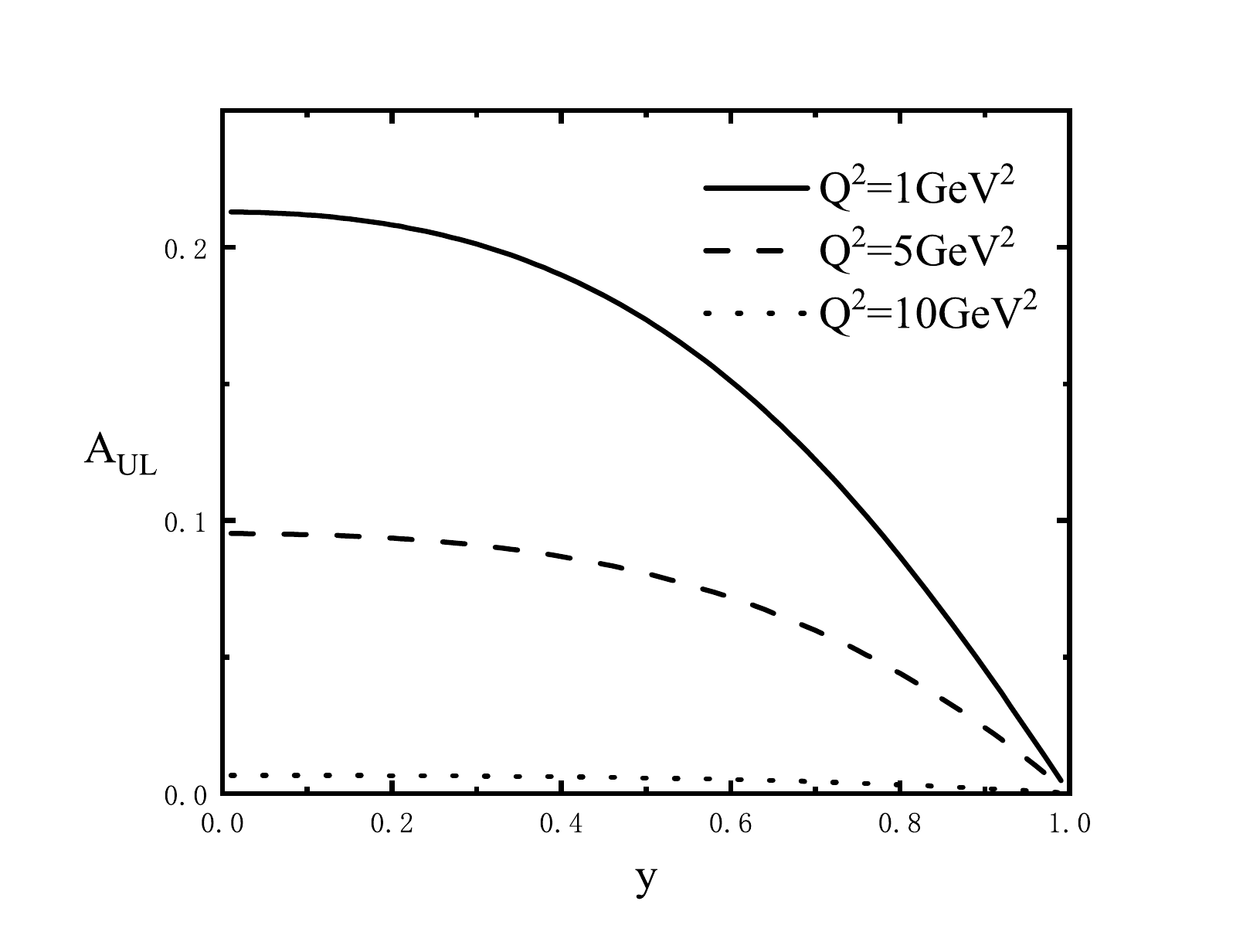}
		\includegraphics[width=80mm]{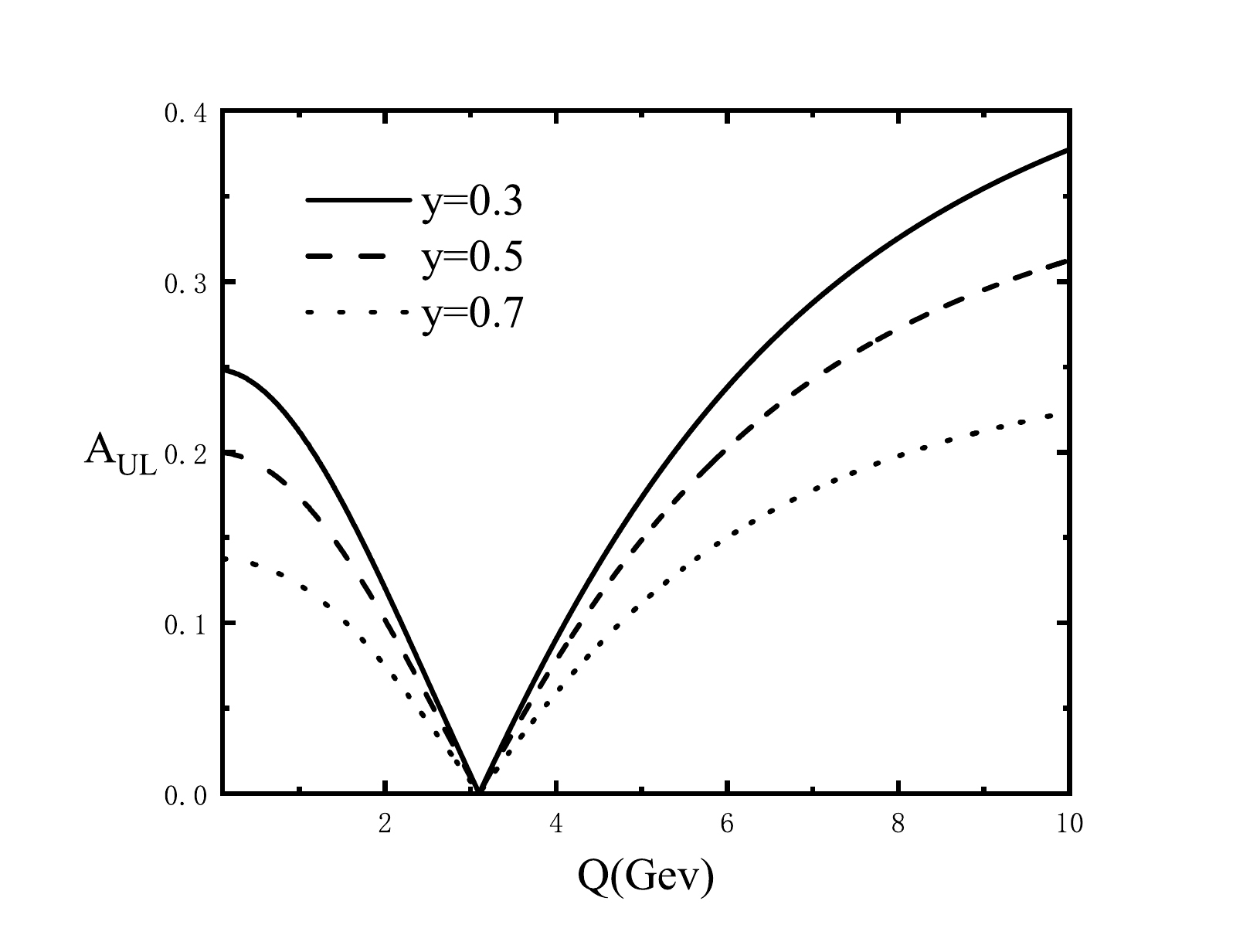}
	\caption{Left: upper bound on the asymmetry $A_{UL}$ as function of $y$ at $Q^2=1,\ 5,\ 10$ GeV$^2$. 
Right:  upper bound on the asymmetry $A_{UL}$ as function of $Q$ at $y=0.3,\ 0.5,\ 0.7$.}\label{fig:AUL}
\end{figure*}

We present the cross sections for this process where the lepton beam or the nucleon target is longitudinally polarized or unpolarized with respect to the direction of its three momentum in the photon-proton center-of-mass frame.
Lepton tensors are either asymmetric or symmetric, and the polarization state of the initial proton can also be considered. 
The polarization of the $J/\psi$ is not considered. 
We use this notation:
\begin{align}
	\frac{d \sigma^{P_{1} P_{2}}}{dy dx_{B} d^2  \boldsymbol{q}_{T}} \equiv d \sigma^{P_{1} P_{2} } (\phi_T)\,,
\end{align}
where the superscript $P_{1}=U/L$ denotes whether leptonic tensor is symmetrical or not, the superscript $P_2=U/T/L$ refers to the possible polarization states of the initial proton.
$d\sigma^{UU}$ and $d\sigma^{UT}$ can be found in Ref.~\cite{Bacchetta:2018ivt}. 
Within these approximations, the final cross-sections read:
\begin{align}\label{eq:csSLU} 
d\sigma^{UL}&=\mathcal{N}|S_{L}|\frac{ \boldsymbol{p}_{T}^2}{M_{p}^2}\ B^{L} h_{1L}^{\perp\,g}(x, \boldsymbol{p}_{T}^2) \ \sin2\phi_{T}\,,\\
\label{eq:csAUU}
d\sigma^{LU}&=0 \,,\\
\label{eq:csATU}
d\sigma^{LT}&=0 \,,\\
\label{eq:csAPU}
d\sigma^{LL}&=\mathcal{N}\ C^{L}\ \Delta g_{1L}(x, \boldsymbol{p}_{T}^2)\,,
\end{align}
where $\phi_{T}$ in Eq.~(\ref{eq:csSLU}) is the azimuthal angle between the transverse momentum of the $J/\psi$ and the lepton plane.
The normalization factor $\mathcal{N}$ is given by
\begin{equation}
   	\mathcal{N}=(2 \pi)^2 \frac{\alpha^2 \alpha_{s} e_{Q}}{y Q^2 M_{\mathcal{Q}} (M_{\mathcal{Q}}^2+Q^2)}\,,
\end{equation}
where $e_{Q}$ is the fractional electric charge of the quark Q. The explicit expression of the terms $B^L,C^L$ can be found in Appendix.

\section{Numerical results}

In this section we present the numerical results for the experimental observables in the $J/\psi$ production in longitudinally polarized DIS.
The mass of $J/\psi$ is set to be $M_{J/\psi}=3.1$ GeV, while the mass of the charm quark is assumed to be $M_{\mathcal{Q}}=M_{J/\psi}/2$.
To obtain numerical estimates for the asymmetries, various sets of extractions of the color-octet LDMEs for $J/\psi$ are available, which are derived from fits to TEVATRON, RHIC, and LHC data.
In this study, we adopt the SV set based on a leading-order calculation~\cite{Sharma:2012dy}.
Since both of the color-octet matrix elements are positive, one has $\langle 0|\mathcal{O}{8}^{J / \psi} (^{1}S{0})|0\rangle=1.8 \times 10^{-2} \mathrm{GeV}^3$ and $\langle 0|\mathcal{O}{8}^{J / \psi} (^{1}P{0})|0\rangle /M_{c}^2=1.8 \times 10^{-2}  \mathrm{GeV}^3$.

\begin{figure*}
	\centering
		\includegraphics[width=80mm]{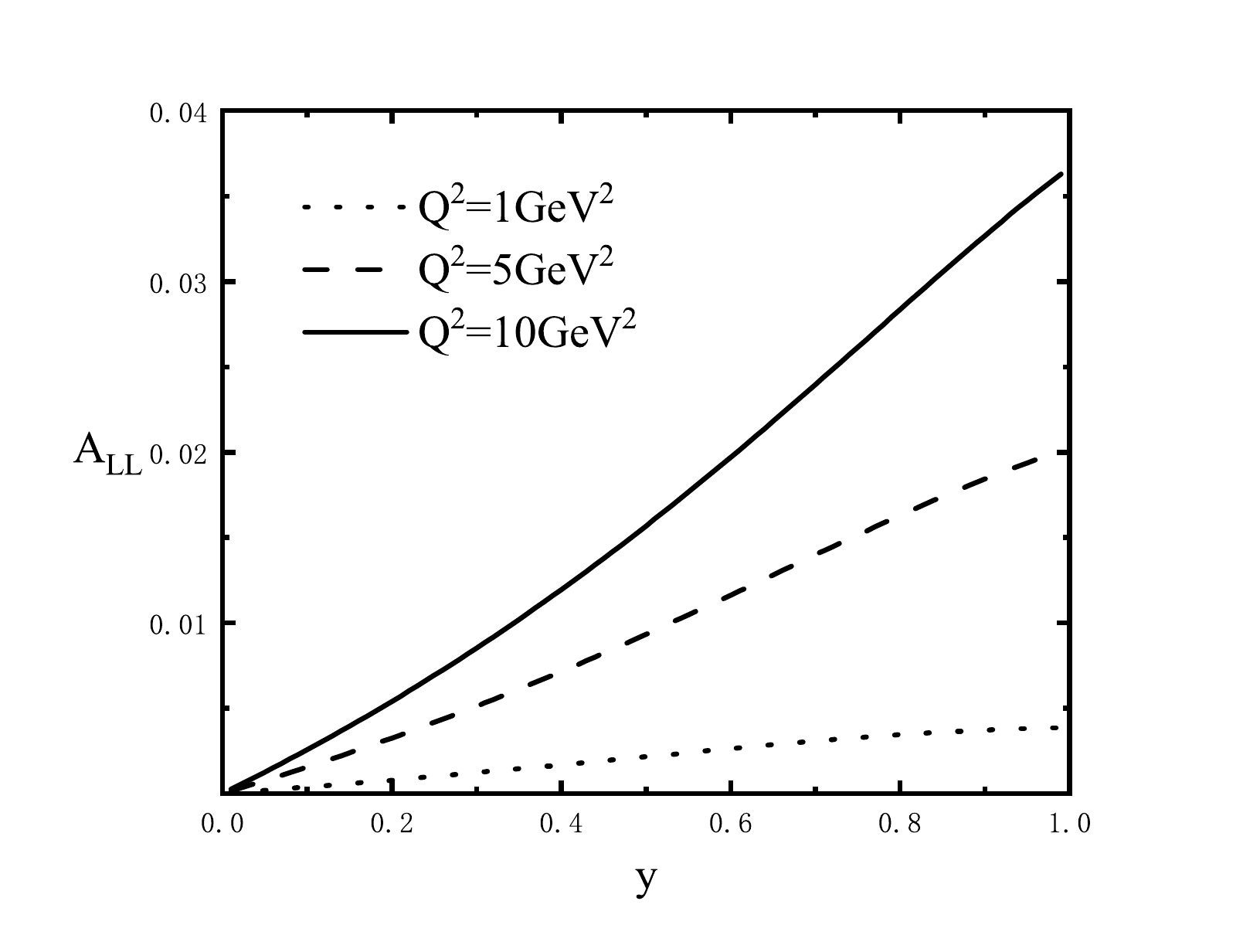}
		\includegraphics[width=80mm]{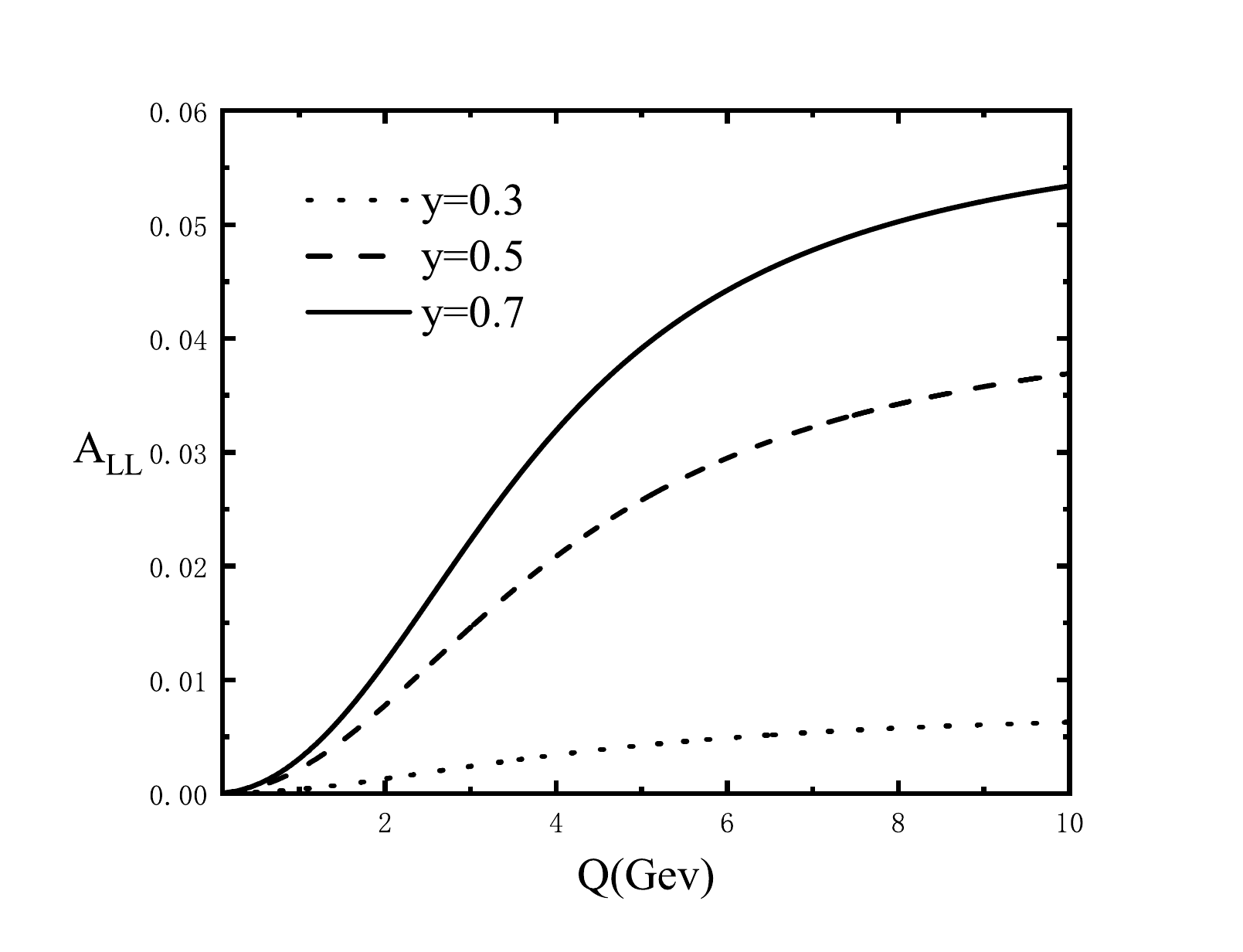}
	\caption{Left: $A_{LL}$ asymmetry in double longitudinally polarized SIDIS as function of $y$ at $Q^2=1$ GeV$^2$, 5 GeV$^2$ and 10 GeV$^2$, respectively. Right:  $A_{LL}$ asymmetry in double longitudinally polarized SIDIS as function of $Q$ at $y=0.3$, 0.5, 0.7, respectively.}
	\label{fig:ALL}
\end{figure*}

The cross section of the process possesses various azimuthal modulations, these modulations can be used to extract information about the ratio of different TMDs~\cite{Bacchetta:2018ivt}.
Therefore, we define the following azimuthal moments in $J/\psi$ production in SIDIS:
\begin{align}\label{eq:azimuthalmoment}
	A^{W(\phi_{T})}\equiv 2 \frac{\int d \phi_{T} W(\phi_{T}) d \sigma (\phi_{T})}{\int d \phi_{T}  d \sigma (\phi_{T})}\,,
\end{align}
where $d \sigma (\phi_{T}) =d \sigma^{UU}+d \sigma^{UL}$.
By taking $W=\sin2\phi_{T}$ we obtain
\begin{align}\label{eq:Asin2} 
	A_{UL}=\left|A^{\mathrm{sin}2\phi_{T}}\right|=\left|\frac{ \boldsymbol{p}_{T}^2}{M_{p}^2}\frac{B^{L}}{A^{U}}\frac{h_{1L}^{\perp\,g}(x, \boldsymbol{p}_{T}^2)}{f_1^g(x, \boldsymbol{p}_{T}^2)}\right|\,.
\end{align}
It is worth noting that the angular structures and TMDs they probe are analogous to the case in transversely polarized proton~\cite{Bacchetta:2018ivt,Chakrabarti:2022rjr}, where $\mathrm{cos}2\phi_{T}$ plays the role of $\mathrm{sin}2\phi_{T}$.

In order to single out $g_{1L}^g(x,p_T^2)$, we define the following azimuthal moment
\begin{align}\label{eq:R}
	A_{LL}= \frac{\int d \phi_{T} d \sigma^{LL}}{\int d d \phi_{T} d \sigma^{UU}} 	=\frac{ C^{L}}{A^{U}}\frac{ g_{1L}^g(x, \boldsymbol{p}_{T}^2)}{f_1^g(x, \boldsymbol{p}_{T}^2)}\,.
\end{align}

For the gluon unpolarized TMD $f_1^g(x, \boldsymbol{p}_{T}^2)$ and the helicity TMD $g_{1L}^g(x,\bm p_T^2)$, we adopt the recent spectator model result from Ref.~\cite{Bacchetta:2020vty}.
As for the T-odd gluon TMD $h_{1L}^{\perp\,g}(x, \boldsymbol{p}_{T}^2)$, 
Since there is no any experimental nor theoretical information, in the calculation we apply the positivity bound in Eq.~(\ref{eq:sharpPositiveBound3}) for it, corresponding to the maximum value of the asymmetry.

In Fig.~\ref{fig:AUL}, we plot the $\sin2\phi_T$ azimuthal asymmetry $A_{UL}$ in the process $e^- + p^\rightarrow \rightarrow e^-+J/\psi + X$. 
The left panel shows the asymmetry as a function of $y$ at $Q^2=1,\ 5,\ 10$ GeV$^2$, while the right one denotes the asymmetry vs $Q$ at $y=0.3,\ 0.5,\ 0.7$.
As mentioned above, this result corresponds to the upper limit of the asymmetry.
The asymmetry has the maximum value at $y=0$ and it decreases with increasing $y$.
On the other hand, the asymmetry vanishes in the region $Q\sim 3.1 $ GeV$^2$, due to the kinematical factor $B_L$ which depends on $Q$. One can also find that this asymmetry is sizable in the smaller and higher $Q$ regions.

In Fig.~\ref{fig:ALL}, we plot the double spin asymmetry $A_{LL}$ in the process ${e^{-}}^\rightarrow+ p^\rightarrow \rightarrow J/\psi + X$ in which the $J/\psi$ are unpolarized.
The left panel and right panels show the asymmetry as a function of $y$ and $Q$, respectively.
We find that in this result the asymmetry is positive since in the spectator model $g_{1L}^g(x,\bm p_T^2)$ is positive.
The asymmetry increases with increasing $y$ or $Q$.

We note that a measurement of the ratio $A^{\sin2\phi_T}/A^{\cos2\phi_T}$ would directly probe the relative magnitude of $ h_{1L}^{\perp,g}/h_{1}^{\perp\,g}$, since the ratio 
\begin{align}\label{eq:ratio}
	\frac{A^{\sin2\phi_T}}{A^{\cos2\phi_T}}=\frac{ h_{1L}^{\perp\,g}(x, \boldsymbol{p}_{T}^2)}{h_{1}^{\perp\,g}(x, \boldsymbol{p}_{T}^2)} \,.
\end{align}
The measurement of the ratio can be used to evaluate the precision on the prediction of gluon TMDs.

Finally, we would like to comment the role of  $h_{1L}^{\perp\,g}$ in azimuthal asymmetries in high-energy processes. 
In our calculation we have not determine the sign of the $\sin 2\phi_T$ asymmetry since we apply the positivity bound for $h_{1L}^{\perp\,g}$. 
It would be interesting to measure the sign of the asymmetry in order to constrain the sign of $h_{1L}^{\perp\,g}$.
Furthermore, as $h_{1L}^{\perp\,g}$  is a T-odd distribution,  it may change the sign between in SIDIS and Drell-Yan due to the gauge-link structure,  analogous to the gluon Sivers function. 
The sign change of $h_{1L}^{\perp\,g}$ in ep collision and pp collision will be of great importance to test the dynamics of QCD.

\section{conclusion}

We studied azimuthal asymmetries in quarkonium production in the SIDIS process with beam or target longitudinally polarized. The observable disentangles the two different gluon TMDs contributing to the longitudinally polarized cross sections, each of them corresponds to a specific azimuthal modulation. 
We calculated the $\sin2\phi_{T}$ asymmetry in $ep$ collision using the TMD formalism and NRQCD. 
This asymmetry is sensitive to the T-odd gluon TMD $h_{1L}^{\perp}(x, \boldsymbol{p}_{T}^{2})$ and is sizeable. 
We present a numerical estimate of the asymmetry in the kinematical region that will be accessible at future EIC using the positivity bound for $h_{1L}^{\perp}(x, \boldsymbol{p}_{T}^{2})$. 
We also estimate the double longitudinal spin asymmetry $A_{LL}$ of $J/\psi$ production using a spectator model result for $g_{1L}^g$.
The measurements of these azimuthal asymmetries would allow a sign change test of the T-odd gluon TMDs by comparing to corresponding observables in hadron-hadron collisions.

\section*{Acknowledgements}
This work is partially supported by the National Natural Science Foundation of China under grant number 12150013.
H. Liu and X. Xie contribute equally to the work and should be considered as the co-first authors.

\section{Appendix}

The explicit expressions of the $\{A,B,C\}$ terms in Eqs.~(\ref{eq:csSLU}),  (\ref{eq:csAPU}) and (\ref{eq:Asin2}) read
  \begin{align}\label{eq:A}
  	A^{U}&=A^{T}=[(1-y)^2+1]\mathcal{A}_{1}^{\gamma*g\rightarrow \mathcal{Q}}-y^2\mathcal{A}_{2}^{\gamma*g\rightarrow \mathcal{Q}}\,,\\
  \label{eq:B}
  	B^{U}&=B^{T}=B^{L}=(1-y)\mathcal{B}^{\gamma*g\rightarrow \mathcal{Q}}\,,\\
\label{eq:C}
  	C^{L}&=[(1-y)^2-1]\mathcal{C}^{\gamma*g\rightarrow \mathcal{Q}}\,.
  \end{align}
In terms of the color-octet LDMEs, one can express
  \begin{align}\label{eq:AU+L_ldmes}
  		\mathcal{A}_{1}^{\gamma*g\rightarrow \mathcal{Q}}=&	\langle 0|\mathcal{O}_{8}^{J / \psi} (^{1}S_{0})|0\rangle  +\frac{4}{N_{c}M_{\mathcal{Q}}^2(M_{\mathcal{Q}}^2+Q^2)^2}\displaybreak[0] \nonumber\\& \times \left[(3M_{\mathcal{Q}}^2+Q^2)^2\langle
  		0|\mathcal{O}_{8}^{J / \psi} (^{3}P_{0})|0\rangle \right.\displaybreak[0] \nonumber \\&+2Q^{2}(2M_{\mathcal{Q}}^2+Q^2)^2\langle
  		0|\mathcal{O}_{8}^{J / \psi} (^{3}P_{1})|0\rangle \displaybreak[0] \nonumber\\&\left.+\frac{2}{5}(6M_{\mathcal{Q}}^4+M_{\mathcal{Q}}^2 Q^2+Q^4)\langle
  		 0|\mathcal{O}_{8}^{J / \psi} (^{3}P_{2})|0\rangle \right] \,,\displaybreak[0]\\
\label{eq:AL_ldmes}
  		\mathcal{A}_{2}^{\gamma*g\rightarrow \mathcal{Q}}=& \frac{16}{N_{c}(M_{\mathcal{Q}}^2+Q^2)^2}
  \nonumber\\& \times \left[ \langle 0|\mathcal{O}_{8}^{J / \psi} (^{3}P_{1})|0\rangle +\frac{3}{5}\langle
  		0|\mathcal{O}_{8}^{J / \psi} (^{3}P_{2})|0\rangle \right]\,,\displaybreak[0]\\
\label{eq:BT_ldmes}
  		\mathcal{B}^{\gamma*g\rightarrow \mathcal{Q}}=&	-\langle 0|\mathcal{O}_{8}^{J / \psi} (^{1}S_{0})|0\rangle  +\frac{4}{N_{c}M_{\mathcal{Q}}^2(M_{\mathcal{Q}}^2+Q^2)^2}\nonumber \displaybreak[0]\\& \times \left[(3M_{\mathcal{Q}}^2+Q^2)^2\langle
  		0|\mathcal{O}_{8}^{J / \psi} (^{3}P_{0})|0\rangle \right. \nonumber\displaybreak[0]\\&-2Q^{4}\langle
  		0|\mathcal{O}_{8}^{J / \psi} (^{3}P_{1})|0\rangle \nonumber\\&\left.+\frac{2}{5}Q^4 \langle
  		0|\mathcal{O}_{8}^{J / \psi} (^{3}P_{2})|0\rangle \right] \,.\displaybreak[0]\\
\label{eq:C_ldmes}
  		\mathcal{C}^{\gamma*g\rightarrow \mathcal{Q}}=&	\langle 0|\mathcal{O}_{8}^{J / \psi} (^{1}S_{0})|0\rangle  +\frac{4}{N_{c}M_{\mathcal{Q}}^2(M_{\mathcal{Q}}^2+Q^2)^2}
  \nonumber \displaybreak[0]\\&
  		\times \left[(3M_{\mathcal{Q}}^2+Q^2)^2\langle
  		0|\mathcal{O}_{8}^{J / \psi} (^{3}P_{0})|0\rangle \right. \nonumber \displaybreak[0]\\&+2 Q^4\langle
  		0|\mathcal{O}_{8}^{J / \psi} (^{3}P_{1})|0\rangle \nonumber\\&\left.+\frac{2}{5}(-6M_{\mathcal{Q}}^4+Q^4)\langle
  		0|\mathcal{O}_{8}^{J / \psi} (^{3}P_{2})|0\rangle \right] \,.
  \end{align}
The expressions can be further simplified by employing the heavy-quark spin symmetry relations.
\begin{align}\label{eq:symmetry relation}
	\langle0|\mathcal{O}_{8}^{J / \psi} (^{3}P_{J})|0\rangle=(2J+1)\langle0|\mathcal{O}_{8}^{J / \psi} (^{3}P_{0})|0\rangle+\mathcal{O}(v^2)\,.
\end{align}
At the leading order in $v$, we can get the following results
\begin{align}
		\mathcal{A}_{1}^{\gamma*g\rightarrow \mathcal{Q}}&=	\langle 0|\mathcal{O}_{8}^{J / \psi} (^{1}S_{0})|0\rangle  \nonumber\displaybreak[0]\\&+\frac{12(7 M_{\mathcal{Q}}^2+3 Q^2)}{N_{c}M_{\mathcal{Q}}^2 (M_{\mathcal{Q}}^2+Q^2)} \langle
		0|\mathcal{O}_{8}^{J / \psi} (^{3}P_{0})|0\rangle\,, \label{eq:AL_ldmesSimple} \displaybreak[0]\\
	\mathcal{A}_{2}^{\gamma*g\rightarrow \mathcal{Q}}&=\frac{96 Q^2}{N_{c}(M_{\mathcal{Q}}^2+Q^2)^2} \langle
		0|\mathcal{O}_{8}^{J / \psi} (^{3}P_{0})|0\rangle \,, 
\label{eq:BT_ldmesSimple}  \displaybreak[0]\\
		\mathcal{B}^{\gamma*g\rightarrow \mathcal{Q}}&=	-\langle 0|\mathcal{O}_{8}^{J / \psi} (^{1}S_{0})|0\rangle  \nonumber\displaybreak[0]\\&+\frac{12(3 M_{\mathcal{Q}}^2- Q^2)}{N_{c}M_{\mathcal{Q}}^2 (M_{\mathcal{Q}}^2+Q^2)} \langle
		0|\mathcal{O}_{8}^{J / \psi} (^{3}P_{0})|0\rangle\,,\label{eq:C_ldmesSimple} \displaybreak[0]\\
		\mathcal{C}^{\gamma*g\rightarrow \mathcal{Q}}=&	\langle 0|\mathcal{O}_{8}^{J / \psi} (^{1}S_{0})|0\rangle  \nonumber\displaybreak[0]\\&+\frac{12(3 Q^2-M_{\mathcal{Q}}^2 )}{N_{c}M_{\mathcal{Q}}^2 (M_{\mathcal{Q}}^2+Q^2)} \langle
		0|\mathcal{O}_{8}^{J / \psi} (^{3}P_{0})|0\rangle\,.
\end{align}

\end{document}